\documentclass{article}

\usepackage[utf8]{inputenc}
\usepackage[T1]{fontenc}
\usepackage{hyperref}
\usepackage{url}
\usepackage{booktabs}
\usepackage{amsfonts}
\usepackage{nicefrac}
\usepackage{microtype}
\usepackage{graphicx}
\usepackage{amsmath}
\usepackage{braket}
\usepackage{float}
\graphicspath{{media/}}
\usepackage{graphicx}
\graphicspath{{Figures/}}
\usepackage{tikz}
\usetikzlibrary{quantikz}
\usepackage{physics}
\usepackage{mathpazo} % Use Palatino font (looks more professional/fancy)
\usepackage[a4paper, margin=1in, top=1.2in, bottom=1.2in]{geometry} % Better margins
\usepackage{xcolor} % For link colors

\hypersetup{
    colorlinks=true,
    linkcolor=blue,
    filecolor=magenta,      
    urlcolor=blue,
    citecolor=blue,
}

\usepackage{fancyhdr}
\makeatletter
\renewcommand{\maketitle}{
  \begin{center}
    \vspace*{0.5cm}
    \hrule height 1.5pt 
    \vspace{0.6cm}
    
    {\Huge \textbf{\textsc{\@title}} \par}
    
    \vspace{0.6cm}
    \hrule height 1.5pt
    \vspace{0.8cm}
    
    {\Large \@author \par}
    \vspace{0.5cm}
  \end{center}
}
\makeatother

\title{An HHL-Based Quantum-Classical Solver for the Incompressible Navier-Stokes Equations with Approximate QST}

\begin{document}

\maketitle

% --- AUTHOR BLOCK (FORCED SIDE-BY-SIDE) ---
    \begin{minipage}[t]{0.45\textwidth}
        \centering
        {Moshe Inger} \\
        Department of Physics \\
        Hebrew University of Jerusalem \\
        Israel \\
        \texttt{moshe.inger@mail.huji.ac.il}
    \end{minipage}
    \hfill % This pushes the two blocks apart
    \begin{minipage}[t]{0.45\textwidth}
        \centering
        {Steven Frankel} \\
        Faculty of Mechanical Engineering \\
        Technion - Israel Institute of Technology \\
        Israel \\
        \texttt{frankel@technion.ac.il}
    \end{minipage}
    
    \vspace{1cm}
    
% --- ABSTRACT (Styled to be narrower) ---
\begin{center}
    \begin{minipage}{0.9\textwidth} % Abstract width = 85% of text width
        \begin{abstract}
            In computational fluid dynamics (CFD), the numerical integration of the Navier–Stokes equations is frequently constrained by the Poisson equation to determine the pressure. Discretization of this equation often results in the need to solve a system of linear algebraic equations. This step typically represents the primary computational bottleneck. Quantum linear system algorithms such as Harrow–Hassidim–Lloyd (HHL) offer the potential for exponential speedups for solving sparse linear systems, such as those that arise from the discretized Poisson equation. In this work, we successfully couple HHL to a discretized formulation of the incompressible Navier–Stokes equations and demonstrate accurate simulations of both lid-driven cavity flow, as a fully integrated benchmark problem, and the Taylor-Green vortex. To address the readout limitation, we utilize a recent novel quantum state tomography (QST) approach based on Chebyshev polynomials and Quantum Amplitude Estimation (QAE), which enables approximate statevector extraction without full state reconstruction. Together, these results clarify the algorithmic structure required for quantum CFD, explicitly confront the measurement bottleneck, and establish benchmark problems for future quantum fluid simulations. We implement the solver using IBM’s Qiskit framework and validate the hybrid quantum-classical simulation against standard classical numerical methods. Our results demonstrate that the hybrid solver successfully captures the global vortex dynamics of the lid-driven cavity problem and the Taylor-Green vortex, offering a robust pathway for integrating quantum subroutines into more practical higher-Reynolds number CFD workflows.
        \end{abstract}
    \end{minipage}
\end{center}

\vspace{1cm} % Space between abstract and intro

% --- KEYWORDS ---
\begin{center}
\end{center}

\section{Introduction}

The numerical solution of the incompressible Navier-Stokes equations is central to modern computational fluid dynamics (CFD), powering applications ranging from aerospace engineering to climate modeling. Standard projection methods decouple the velocity and pressure fields, necessitating the solution of the Poisson equation to determine the pressure at every discrete time step. This constitutes the primary computational bottleneck, consuming up to 90\% of the total execution time due to the size and conditioning of the resulting linear systems \cite{Ozbay2021}.

Quantum algorithms for solving linear systems, most notably the Harrow-Hassidim-Lloyd (HHL) algorithm \cite{hhl09}, offer a theoretical exponential advantage in solving sparse linear systems under specific conditions \cite{clader2013preconditioned, childs2017quantum}. Although practical quantum advantage for industrial-scale CFD remains an open challenge, hybrid quantum-classical architectures provide a promising pathway for the near-term integration of quantum subroutines. Previous work has already explored using the quantum amplitude estimation algorithm \cite{gaitan2020finding}, the approximate quantum Fourier transform \cite{steijl2019quantum}, the variational quantum linear solver (VQLS) algorithm \cite{song2025incompressible}, lattice-Boltzmann methods (LBM) \cite{zeng2025quantum}, and even HHL \cite{lapworth2022implicit}.

Established benchmarks have also already validated using HHL to solve the finite-difference approximation of the Poisson equation \cite{ghafourpour25}. In this work, we build on these results, developing and validating an end-to-end iterative quantum-classical Navier-Stokes solver. Our framework incorporates the HHL algorithm to solve the Poisson equation within a classical projection scheme. Since this is a hybrid quantum-classical algorithm, we are primarily motivated by the theoretically predicted exponential speedup of HHL on actual fault-tolerant hardware. We still have the advantage of encoding $2^n$ data points in $n$ problem qubits while solving it, but since those data points are passed to and from the classical computer, we still need $2^n$ classical bits, forgoing the advantage provided by the larger Hilbert space qubits inhabit. 

As Scott Aaronson highlights, HHL only prepares a quantum state proportional to the solution, in our case, $\ket p$, the pressure \cite{aaronson2015read}. Many quantum computing algorithms manipulate data encoded in the amplitudes of a superposition of qubits. By the Born rule of quantum mechanics, these amplitudes correspond to the probability of finding the quantum system in that particular state. Measuring the system simply collapses the superposition of various amplitudes, our data, into a single state according to the states' probabilities. One of the greatest outstanding problems in quantum computing is the readout problem, accessing the data encoded in the amplitudes without requiring an exponential amount of measurements. Fully recreating the statevector to determine each amplitude scales exponentially with the number of qubits \cite{Brandao2021}. To address the readout problem, we implemented Chebyshev-based Quantum State Tomography (QST), introduced by Su et al. \cite{su2025efficient}. We replace the aforementioned traditional statevector readout methods with a spectral projection method. By projecting the quantum state onto a truncated basis of $m$ Chebyshev polynomials, we achieve a compressed representation of the pressure field. In the case of the lid-driven cavity, we employ a non-uniform, hyperbolic stretching function to map the computational grid, allowing for the resolution of high-gradient boundary layers, as described in subsection~\ref{subsec:chebyshev}. Su et al. validated the potential for this method to capture the dominant modes of the encoded quantum state, and the efficacy of a curvilinear grid to improve accuracy \cite{su2025efficient}. 

The final challenge of QST is determining the correct magnitude of the resultant statevector, since all quantum statevectors are normalized. To address this, we utilize Quantum Amplitude Estimation (QAE), a quantum algorithm inspired by Grover's search algorithm \cite{grover1997quantum}, which, like Grover's algorithm, provides a theoretical quadratic speedup over classical statistical sampling methods \cite{brassard2000quantum}. 

\noindent We implement this solver via IBM's Qiskit framework \cite{qiskit}.

\section{Methodology}
\subsection{Classical Algorithm}
\subsubsection{Governing Equations and Projection Method \label{subsec:equations}}
We consider the incompressible Navier-Stokes equations:
\begin{align}
\frac{\partial \mathbf{u}}{\partial t} + (\mathbf{u} \cdot \nabla)\mathbf{u} &= -\frac{1}{\rho}\nabla p + \nu \nabla^2 \mathbf{u} \\
\nabla \cdot \mathbf{u} &= 0
\end{align}

\noindent where $\textbf{u}$ is the velocity, $p$ is the pressure, $\nu$ is the kinematic viscosity ($\frac{1}{Re}$), and $\rho$ is the mass density. We employ a standard split-step projection scheme \cite{chorin68}. First, an intermediate velocity $\mathbf{u}^*$ is computed explicitly, on a classical computer, by decoupling the pressure field:
\begin{equation}
\frac{\mathbf{u}^* - \mathbf{u}^n}{\Delta t} = - (\mathbf{u}^n \cdot \nabla)\mathbf{u}^n + \nu \nabla^2 \mathbf{u}^n
\end{equation}

\noindent Taking the divergence of the pressure correction step yields the Poisson equation:
\begin{equation}
\nabla^2 p = \frac{\rho}{\Delta t} \nabla \cdot \mathbf{u}^*
\end{equation}

\noindent We use our quantum subroutine to determine $p$, which we then return to the classical computer to project the velocity field onto the divergence-free subspace to enforce incompressibility at the next time step:
\begin{equation}
\mathbf{u}^{n+1} = \mathbf{u}^* - \frac{\Delta t}{\rho} \nabla p
\end{equation}

\subsubsection{Numerical Discretization \label{subsec:numerical_discretization}}
The numerical framework relies on a decoupled approach for temporal and spatial discretizations. Advancement in time is treated explicitly using a first-order Forward Euler scheme \cite{anderson95}. The continuous time derivative for any given flow variable $\phi$ is approximated as:
\begin{equation}
\frac{\partial \phi}{\partial t} \approx \frac{\phi^{n+1} - \phi^n}{\Delta t} + \mathcal{O}(\Delta t)
\end{equation}
where the superscript $n$ denotes the solution evaluated at time $t = n \Delta t$, and $\Delta t$ is the uniform discrete time step size.

Spatially, we first discretize the 2D domain $[0,L]^2$ onto a two-dimensional Cartesian grid indexed by $(i,j)$. Spatial derivatives are approximated using second-order central differences, scaled by the local metric coefficients $h_{\xi,i}$ and $h_{\eta,j}$ to map between the physical and computational spaces $(\xi, \eta)$. Combining the Forward Euler temporal discretization with the spatial central difference operators, the intermediate velocity field $\mathbf{u}^* = (u^*, v^*)$ is advanced as follows:
\begin{align}
u^*_{i,j} &= u_{i,j}^n + \Delta t \left[ - \left( u_{i,j}^n h_{\xi,i} \frac{u_{i+1,j}^n - u_{i-1,j}^n}{2\Delta \xi} + v_{i,j}^n h_{\eta,j} \frac{u_{i,j+1}^n - u_{i,j-1}^n}{2\Delta \eta} \right) + \nu \nabla^2 u_{i,j}^n \right] \\
v^*_{i,j} &= v_{i,j}^n + \Delta t \left[ - \left( u_{i,j}^n h_{\xi,i} \frac{v_{i+1,j}^n - v_{i-1,j}^n}{2\Delta \xi} + v_{i,j}^n h_{\eta,j} \frac{v_{i,j+1}^n - v_{i,j-1}^n}{2\Delta \eta} \right) + \nu \nabla^2 v_{i,j}^n \right]
\end{align}
where the discrete Laplacian operators $\nabla^2 u_{i,j}^n$ and $\nabla^2 v_{i,j}^n$ utilize the standard 2D 5-point spatial finite difference method. For the scalar pressure field ($p$), the discrete two-dimensional gradient operator $\nabla p$ at node $(i,j)$ is approximated as:
\begin{equation}
\nabla p_{i,j} \approx \left( 
h_{\xi,i} \frac{p_{i+1,j} - p_{i-1,j}}{2\Delta \xi}, \;\; 
h_{\eta,j} \frac{p_{i,j+1} - p_{i,j-1}}{2\Delta \eta} 
\right)
\end{equation}

\noindent The second-order Laplacian operator $\nabla^2$ is discretized using a standard 5-point stencil. The discrete 2D Laplacian applied to the pressure field is expressed as:
\begin{equation}
\nabla^2 p_{i,j} \approx h_{\xi,i}^2 \frac{p_{i+1,j} - 2p_{i,j} + p_{i-1,j}}{\Delta \xi^2} + h_{\eta,j}^2 \frac{p_{i,j+1} - 2p_{i,j} + p_{i,j-1}}{\Delta \eta^2}
\end{equation}

Solving this Poisson equation is the primary computational bottleneck in classical CFD. Therefore, this work aims to solve it via our quantum subroutine (as described in subsection~\ref{subsec:quantum_interface}). To accomplish this, the two-dimensional spatial grid must be mapped to a one-dimensional vector. This is done by mapping the 2D indices $(i,j)$ to a 1D vector index $k = i + (j-1)N_{\xi}$, where $N_{\xi}$ and $N_{\eta}$ are the total number of grid points in the $\xi$ and $\eta$ directions, respectively. This ordering translates the 5-point spatial stencil into a sparse, block-tridiagonal matrix structure for $A$:
\begin{equation}
A = 
\begin{pmatrix}
D_1 & U_1 & 0 & \cdots & 0 \\
L_2 & D_2 & U_2 & \ddots & \vdots \\
0 & \ddots & \ddots & \ddots & 0 \\
\vdots & \ddots & L_{N_{\eta}-1} & D_{N_{\eta}-1} & U_{N_{\eta}-1} \\
0 & \cdots & 0 & L_{N_{\eta}} & D_{N_{\eta}}
\end{pmatrix}
\end{equation}

\noindent Here, each diagonal block $D_j$ is an $N_{\xi} \times N_{\xi}$ tridiagonal matrix capturing the intra-row spatial couplings along the $\xi$-direction:
\begin{equation}
D_j = 
\begin{pmatrix}
c_{1,j} & a_{1,j} & 0 & \cdots & 0 \\
a_{2,j} & c_{2,j} & a_{3,j} & \ddots & \vdots \\
0 & \ddots & \ddots & \ddots & 0 \\
\vdots & \ddots & a_{N_{\xi}-1,j} & c_{N_{\xi}-1,j} & a_{N_{\xi},j} \\
0 & \cdots & 0 & a_{N_{\xi},j} & c_{N_{\xi},j}
\end{pmatrix}
\end{equation}

\noindent The off-diagonal blocks $L_j$ and $U_j$ are $N_{\xi} \times N_{\xi}$ diagonal matrices capturing the inter-row couplings along the $\eta$-direction:
\begin{equation}
L_j = U_j = \text{diag}\left( b_{1,j}, b_{2,j}, \dots, b_{N_{\xi},j} \right)
\end{equation}

\noindent Matching the discrete 2D Laplacian operator, the non-zero coefficients for any interior node are entirely determined by the local grid metric terms:
\begin{align}
c_{i,j} &= -2 \left( \frac{h_{\xi,i}^2}{\Delta \xi^2} + \frac{h_{\eta,j}^2}{\Delta \eta^2} \right) \quad &\text{(Main diagonal)} \\
a_{i,j} &= \frac{h_{\xi,i}^2}{\Delta \xi^2} \quad &\text{($\xi$-direction coupling)} \\
b_{i,j} &= \frac{h_{\eta,j}^2}{\Delta \eta^2} \quad &\text{($\eta$-direction coupling)}
\end{align}

\subsection{Quantum Algorithm}
\subsubsection{Quantum-Classical Hybrid Interface (HHL) \label{subsec:quantum_interface}}

\begin{figure}[H]
    \centering
    \begin{tikzcd}
    \ket{0}^{\otimes n_b} \qw & \gate{\textbf{State Prep}} & \gate[2]{\textbf{QPE }} & \qw & \gate[2]{\textbf{QPE}^\dagger } & \qw & \ket{x}_{n_b} \\
    \ket{0}^{\otimes n_l} \qw & \qw & \ctrl{-1} & \ctrl{1} & \ctrl{-1} & \qw & \ket{0}^{\otimes n_l} \\
    \ket{0} \qw & \qw & \qw & \gate{\textbf{R}_y} & \qw & \qw & \sqrt{1-\frac{C^2}{\lambda^2}}\ket{0} + \frac{C}{\lambda}\ket{1}
    \end{tikzcd}
    \caption{Quantum Circuit Diagram of HHL. \textbf{Top Row}: The "b register" where the right-hand side of $A\mathbf{p}=\mathbf{b}$ is encoded, and the final state, $\ket{x} \propto A^{-1}\mathbf{b}$ is found at the end. \textbf{Middle Row}: The "clock register" where the eigenvalues of $A$ are estimated and subsequently inverted. \textbf{Bottom Row}: The ancilla qubit, used to determine if the entangled b register and clock register collapsed to the solution state. Architecture adapted from Ghafourpour and Laizet \cite{ghafourpour25}.}
\label{fig:1}
\end{figure}
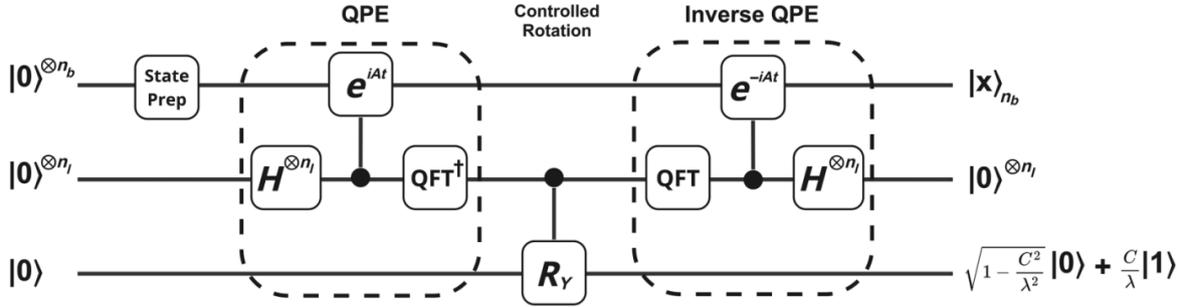

\textbf{The HHL Algorithm:}

\noindent HHL is a fundamental quantum procedure designed to solve the linear system $A\mathbf{x} = \mathbf{b}$ by preparing a quantum state $\ket{x} = A^{-1}\ket{b}$. HHL prepares this quantum state by encoding the eigenvalues of $A$ as phases in a quantum register, allowing them to be inverted much more efficiently than through explicit calculation. In particular, the algorithm executes through three primary stages, displayed in Figure \ref{fig:1}:

First, the input vector $\mathbf{b}$ is encoded as a quantum state $\ket{b} = \sum_{j} \beta_j \ket{u_j}$ in the b register, where $\ket{u_j}$ are the eigenvectors of $A$. Quantum Phase Estimation (QPE) \cite{Cleve1998} is then applied, utilizing Hamiltonian simulation $e^{iAt}$. This step extracts the eigenvalues $\lambda_j$ of $A$ and stores them in the clock register, resulting in the entangled state $\sum_{j} \beta_j \ket{\lambda_j}\ket{u_j}$. For this reason, HHL is sensitive to the range between the minimum and maximum eigenvalues, i.e, the conditioning of the matrix $A$. This is important, as the conditioning of $A$ will depend on $\beta$, the stretching factor of the curvilinear coordinates. 

Second, an eigenvalue inversion is performed using a controlled rotation on an ancilla qubit, a workspace qubit to ensure that at the end of the algorithm the entangled quantum state has collapsed correctly. This operation is conditioned on the clock register and transforms the state of the ancilla qubit based on the reciprocal of the stored eigenvalues. The rotation maps the value $\frac{1}{\lambda_j}$ to the amplitude of the $\ket{1}$ state of the ancilla qubit, yielding a state proportional to $\sum_{j} \beta_j \ket{\lambda_j}\ket{u_j} \left( \frac{C}{\lambda_j} \ket{1} + \sqrt{1 - \frac{C^2}{\lambda_j^2}} \ket{0} \right)$, where $C$ is a normalization constant.

Finally, the inverse Quantum Phase Estimation (QPE$^\dagger$) is executed to uncompute the clock register, returning it to the $\ket{0}$ state. Upon measuring the ancilla qubit and successfully obtaining the outcome $\ket{1}$, the b register collapses into the solution state $\ket{x} \propto \sum_{j} \beta_j \frac{1}{\lambda_j} \ket{u_j}$, proportional to $A^{-1}\ket{b}$. Since probability must always be a number between 0 and 1, and the amplitudes correspond to the probability of measuring that particular quantum state, the constant $C$ will depend on the conditioning of the matrix $A$; the greater the range between the minimum and maximum eigenvalues, the lower $C$ will be, making it potentially more difficult to measure $\ket{1}$ on the ancilla qubit. This is one of the known limitations of HHL. 

\noindent\textbf{Quantum Subroutine:}

\noindent The core of our solver is a hybrid interface that leverages the HHL algorithm to solve the discretized Poisson equation, as follows:

For this benchmark, we assume access to an ideal state preparation oracle that loads the normalized velocity divergence vector $|b\rangle$, the right hand side of $A\mathbf{p}=\mathbf{b}$ described in subsection \ref{subsec:numerical_discretization}, into the quantum register with amplitude encoding. This is one of the biggest open challenges in quantum computing. The cost of this preparation is not included in the runtime analysis.

The matrix $A$ was decomposed into a sum of Pauli strings $A = \sum c_k P_k$ using Qiskit's \textsf{SparsePauliOp} decomposition. The time-evolution operator $U = e^{i A t}$ was approximated using a first-order Lie-Trotter product formula with $R=150$ Trotter steps to suppress discretization error \cite{lloyd1996universal}.

The eigenvalue inversion step was performed using coherent arithmetic inversion via Qiskit's \textsf{ExactReciprocal} function, which computes the reciprocal of the phase estimation register into an ancillary register before performing the controlled rotation.

Rather than simple statevector extraction, we perform a spectral reconstruction. The quantum state $\ket{p}$ is projected onto a basis of $m$ Chebyshev polynomials $T_k(\xi)$, defined and described in subsection~\ref{subsec:chebyshev}, via the Hadamard test, a foundational quantum algorithm used to determine the overlap between an encoded quantum state, $\ket{\psi}$, and a unitary $U$ \cite{Aharonov1998}. To limit the runtime of our simulations, we emulated the readout process classically. We computed exact inner products and introduced binomial shot noise corresponding to $N_{shots} = 10^7$ to rigorously model the statistical uncertainty of the quantum measurement. Finally, a zero-mean gauge condition $\int p \, d\Omega = 0$ is enforced to ensure a unique solution for the all-Neumann system.

 \subsubsection{Chebyshev Polynomials and Curvilinear Mapping \label{subsec:chebyshev}}

We estimate the encoded quantum state, $\ket{p}$, by projecting it onto a truncated basis of $m$ Chebyshev polynomials. Chebyshev polynomials are defined as:
\begin{equation}
    T_{i+1}(x) = 2xT_i(x) - T_{i-1}(x), \quad i \ge 1
\end{equation}

\noindent beginning with $T_0(x) = 1$ and $T_1(x) = x$. Chebyshev polynomials are orthogonal over the correct coordinate system, and always have orthogonal components in any coordinate system. Therefore, they serve as an adequate basis to construct any function, provided the Gram correction matrix is accounted for when calculating the coefficient for each polynomial. Truncating the basis to a finite number of polynomials has been shown to adequately capture the dominant features of continuous functions and even collections of data points \cite{su2025efficient}.

To improve the accuracy of the projection onto Chebyshev polynomials, we employ a curvilinear coordinate transformation. We map the uniform computational space $\boldsymbol{\xi} = (\xi, \eta) \in [0, 1]^2$ to the physical domain $\mathbf{x} = (x, y)$ using a hyperbolic stretching function:
\begin{equation}
x_i = \frac{L}{2} \left[ 1 + \beta \tan^{-1} \left( (2\xi_i - 1) \tan(1/\beta) \right) \right]
\end{equation}
where $\beta$ is a stretching parameter that clusters grid points near the boundaries. We use $\beta = 2.5$ in our simulations. The equations are solved in the computational domain by transforming the differential operators using the metric Jacobian $h_i = \frac{\partial \xi_i} {\partial x_i}$. To ensure that the Laplacian matrix $A$ remains Hermitian on the non-uniform grid, we employed a symmetric metric discretization. The off-diagonal coupling terms were scaled by the product of adjacent metric coefficients, ensuring $A_{ij} = A_{ji}$.

\subsubsection{Determining Scaling Factor through Quantum Amplitude Estimation}

While QST successfully extracts the geometric shape of the pressure field by reconstructing a normalized statevector, it is unable to determine the appropriate magnitude. To bridge this gap without relying on classical reference data, we employ Quantum Amplitude Estimation (QAE) \cite{brassard2000quantum}, which provides a theoretical quadratic speedup over classical Monte Carlo sampling for evaluating the probability amplitude of a designated quantum state, as follows:

\noindent Given a state preparation operator $\mathcal{A}$ that acts on an initial state $|0\rangle$, QAE estimates the probability $a$ of measuring a specific target state $|\psi_{\text{target}}\rangle$:

\begin{equation}
a = |\langle \psi_{\text{target}} | \mathcal{A} | 0 \rangle|^2
\end{equation}

This probability could be approximated via repeated binomial sampling. Rather than relying on this, QAE combines the mechanics of Grover's search algorithm \cite{grover1997quantum} with QPE. The algorithm defines a unitary Grover iteration operator, $\mathcal{Q}$:

\begin{equation}
\mathcal{Q} = -\mathcal{A} \mathcal{S}_0 \mathcal{A}^{-1} \mathcal{S}_{\psi}
\end{equation}

\noindent where $\mathcal{S}_0$ and $\mathcal{S}_{\psi}$ are reflection operators about the initial state and the target state, respectively. The eigenvalues of $\mathcal{Q}$ take the form $e^{\pm i 2\theta}$, which inherently encode the target probability via the relation $a = \sin^2(\theta)$. By applying QPE to the operator $\mathcal{Q}$, the phase $\theta$ can be extracted directly. This is achieved using an auxiliary register prepared in a superposition by a Quantum Fourier Transform. This register controls successive applications of $\mathcal{Q}^j$ on the target state. Finally, an inverse Quantum Fourier Transform is applied to the auxiliary register to extract the phase. The precision of this estimation scales as $\epsilon \sim \mathcal{O}(\frac{1}{M})$, where $M$ represents the dimension of the auxiliary state space and the maximum number of coherent applications of $\mathcal{Q}$. In other words, to achieve an accuracy $\epsilon$, QAE must be applied $\mathcal{O}(M)$ times. This establishes a strict quadratic advantage over the $\mathcal{O}(\frac{1}{\epsilon^2})$ convergence limit of classical sampling methods, where, to achieve an accuracy $\epsilon$, one must sample $\mathcal{O}(M^2)$ times. 

In the HHL algorithm, the probability of measuring the ancilla flag qubit in the $\ket{1}$ state, denoted $P(1)$, is proportional to the squared norm of the unnormalized quantum solution. By applying QAE, we obtain a bounded estimate of $P(1)$ with a target accuracy $\epsilon$. The true physical magnitude of the pressure field, $||\vec{x}_{\text{true}}||$, is then deterministically recovered using the relation:

\begin{equation}
||\vec{x}_{\text{true}}|| = \frac{\sqrt{P(1)}}{C} ||\vec{b}|| 
\end{equation}

\noindent where $C$ is the reciprocal scaling constant from the controlled eigenvalue inversion and $||\vec{b}||$ is the physical norm of the classical input divergence field. Finally, we multiply all the Chebyshev QST coefficients by this scaling factor. This is a standard method of extracting a single quantum amplitude, even for the specific purpose of scaling the resultant statevector of HHL \cite{lin2022lecture}.

Lastly, another relevant postulate of quantum mechanics is that all quantum states are equivalent up to a global phase. In other words, $\ket{p}$ and $-\ket{p}$ represent the same physical reality when encoded in qubits. Therefore, after our QST procedure, we compute the inner product $(A \vec{x}_{\text{qst}}) \cdot \vec{b}$ to verify the correct direction. If the dot product is negative, meaning the vectors are misaligned, the QST procedure determined the state with an inverted phase, which we simply correct by multiplying by -1. This operation incurs a computational cost of $\mathcal{O}(N)$, where $N$ is the total number of spatial grid points, which we already incur per time step simply by extracting the field, calculating the local pressure gradients, and updating the discrete classical velocity grid. So, in the context of a hybrid quantum-classical solver, this operation incurs no additional computational complexity.

\section{Results}

\subsection{Solving the Poisson Equation}
\begin{figure}[H]
\centering
\includegraphics[width=1.0\textwidth]{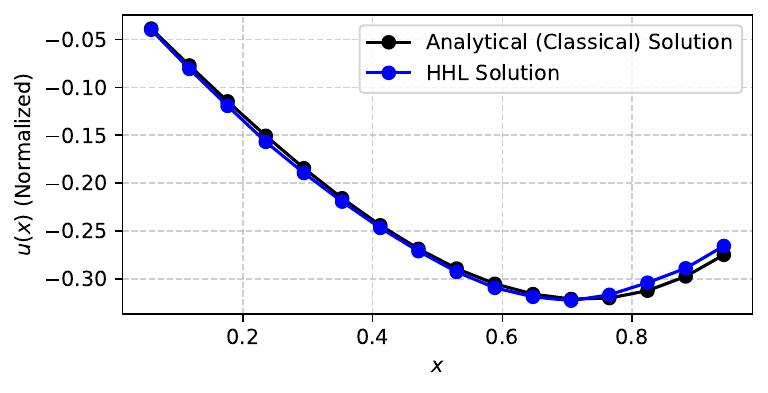}
\caption{HHL Solution to the 1D Poisson Equation, $\frac{d^2u}{dx^2}=f(x), f(x) =10x$, domain $[0,1]$, inhomogeneous boundary conditions: $u(0)=0, u(1) = 1$. 16 grid points using $n=4$ problem qubits, $n_c = 8$ clock qubits, and $R=150$ Trotter steps.}
\label{fig:3}
\end{figure}

\begin{figure}[h]
\centering
\includegraphics[width=1.0\textwidth]{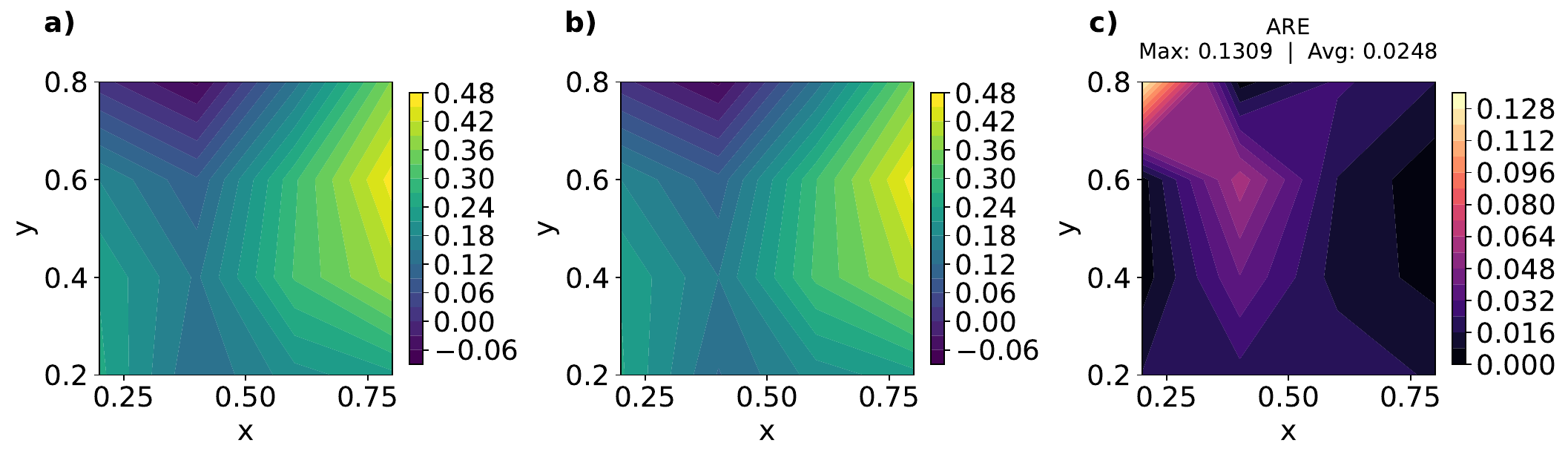}
\caption{a) Solution to the 2D Poisson Equation, $\nabla^2u=f(x,y), f(x,y) =4- 8H(x-0.5)$, domain $[0,1]^2$, inhomogeneous boundary conditions: $u(0, y) = 0.5, u(1, y) = \sin(y), u(x, 0) = (x - 0.5)(x - 1.0), u(x, 1) = 0.5(x - 1.0)$. a) Classical Benchmark. b) HHL Solution, 16$\times$16 grid using $n=8$ problem qubits, $n_c = 8$ clock qubits, and $R=150$ Trotter steps. c) ARE between Classical and HHL Solution.}
\label{fig:4}
\end{figure}

Firstly, we reproduced previous results using HHL to solve the discretized Poisson equation, in both one and two dimensions. Figure \ref{fig:3} displays the results for the 1D case, and figure \ref{fig:4} displays the results for the 2D case, both following the benchmark established by Ghafourpour and Laizet \cite{ghafourpour25}. For the 1D case, we solve $\frac{d^2u}{dx^2}=f(x), f(x) =10x$, over the domain $[0,1]$, with inhomogeneous boundary conditions $u(0)=0, u(1) = 1$. We use 16 grid points, encoded in $n=4$ problem qubits, with $n_c = 8$ clock qubits, and $R=150$ Trotter steps.

\noindent For the 2D case, we solve the Poisson equation $\nabla^2u(x,y) =4- 8H(x-0.5)$ over the domain $[0,1]^2$, with inhomogeneous boundary conditions:
\begin{align*}
\text{Left Boundary: }   & u(0, y) = 0.5, & 0 \leq y \leq 1 \\
\text{Right Boundary: }  & u(1, y) = \sin(y), & 0 \leq y \leq 1 \\
\text{Bottom Boundary: } & u(x, 0) = (x - 0.5)(x - 1.0), & 0 \leq x \leq 1 \\
\text{Top Boundary: }    & u(x, 1) = 0.5(x - 1.0), & 0 \leq x \leq 1
\end{align*}

\noindent We measure the results qualitatively using Absolute Relative Error (ARE): 

\begin{equation}
    \text{ARE} = \frac{||\mathbf{u}_{hybrid}| - |\mathbf{u}_{classical}||}{|\mathbf{u}_{classical}| + \epsilon}
\end{equation}

\noindent The results are as accurate as conventional classical methods, with an average ARE of only 0.02280 ($\approx 2.3\%$) for the 1D case, and 0.02475 ($\approx2.5\%$) for the 2D case. 

\begin{figure}[H]
\centering
\includegraphics[width=1.0\textwidth]{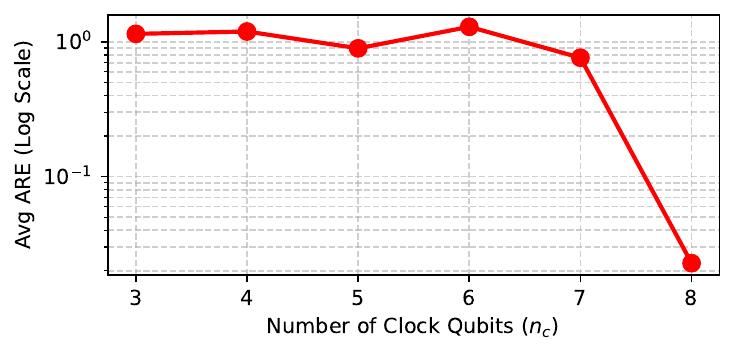}
\caption{Average ARE of HHL Poisson Solver to the 1D Poisson Equation, $\frac{d^2u}{dx^2}=f(x), f(x) =10x$, domain $[0,1]$, inhomogenous boundary conditions: $u(0)=0, u(1) = 1$. 16 grid points using $n=4$ problem qubits, as a Function of $n_c$, the number of clock qubits.}
\label{fig:5}
\end{figure}

To further investigate this use of HHL, we plotted the average ARE of our HHL solution to the 1D Poisson Equation, $\frac{d^2u}{dx^2}=f(x), f(x) =10x$, over 16 grid points, as a function of $n_c$, the number of clock qubits. The results are displayed in Figure \ref{fig:5}. 
Firstly, we note the erratic change in accuracy, as opposed to the expected steady decrease of ARE as $n_c$ increases. Additionally, we note the sharp decrease of ARE at $n_c =8$, the number of clock qubits we used for the remainder of our simulations.  

\subsection{Full State Reconstruction to Solve Navier-Stokes}
\begin{figure}[H]
\centering
\includegraphics[width=1.0\textwidth]{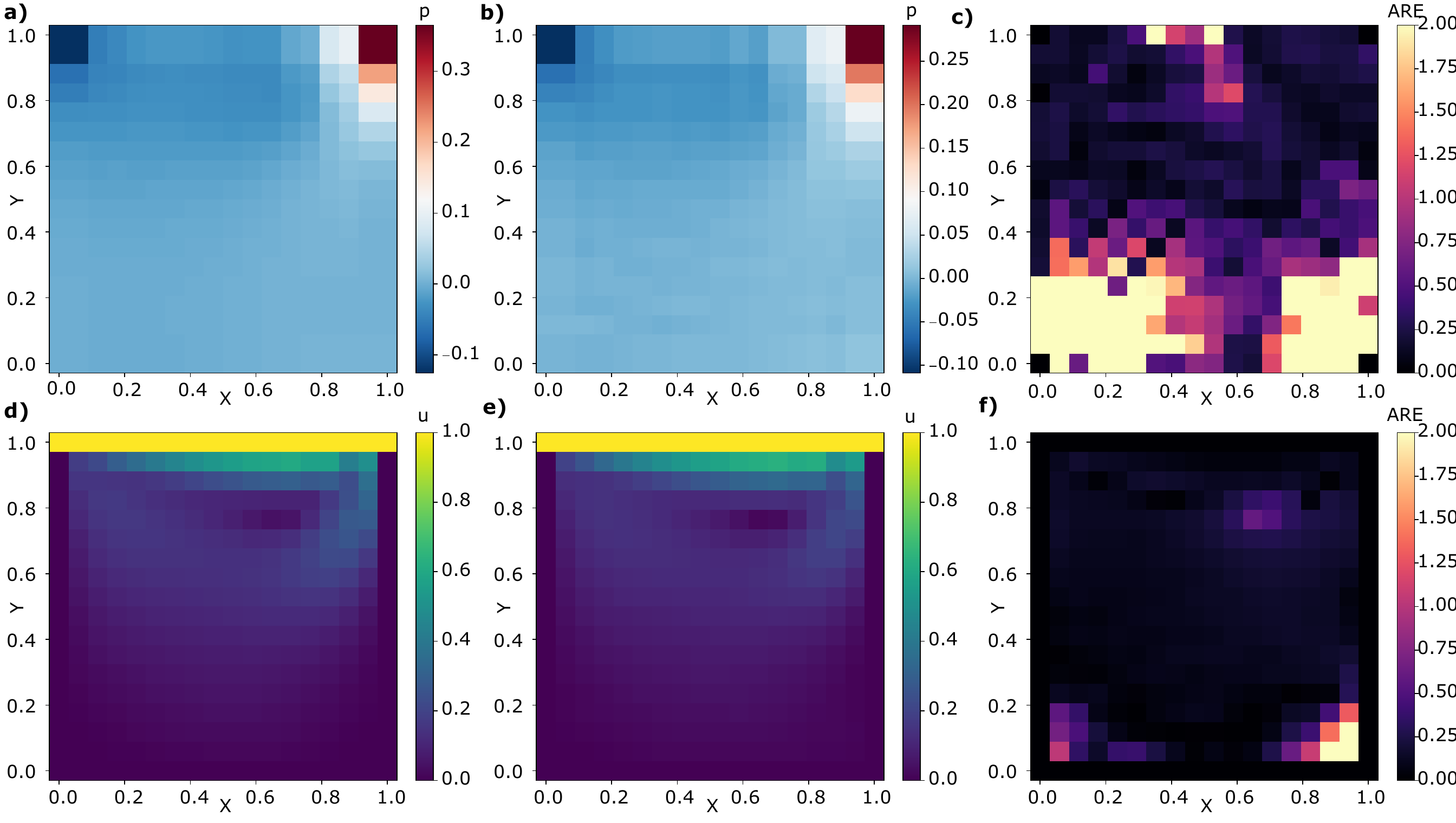}
\caption{Comparison of 2D Lid-Driven Cavity flow with Full State Vector Extraction. $Re=100$, $16 \times 16$ grid, $n_c=8$ clock qubits, $R=150$ Trotter steps. Pressure Plot: a) Classical Benchmark. b) HHL Quantum-Classical Hybrid Solution. Velocity Plot: d) Classical Benchmark. e) Quantum-Classical Hybrid Solution. ARE: c) of Pressure Gradient. f) of Velocity.}
\label{fig:6}
\end{figure}

We then inserted the HHL solver for the 2D Poisson equation into an otherwise classical Navier-Stokes solver (as described in subsection \ref{subsec:equations} above), using the output of HHL to update the next time step of the velocity. At this stage, we used Qiskit's \textsf{getstatevector()} function to access the entire resultant quantum state, which we could not do on actual quantum hardware. The average ARE of the velocity field is only 16\%, which is primarily located at the bottom of the cavity where the velocity is close to 0, amplifying the actual error. As Figure \ref{fig:6} displays, the error is close to 0\% in a majority of the cavity.

\subsection{Quantum State Tomography via Chebyshev Polynomials}
\begin{figure}[H]
\centering
\includegraphics[width=1.0\textwidth]{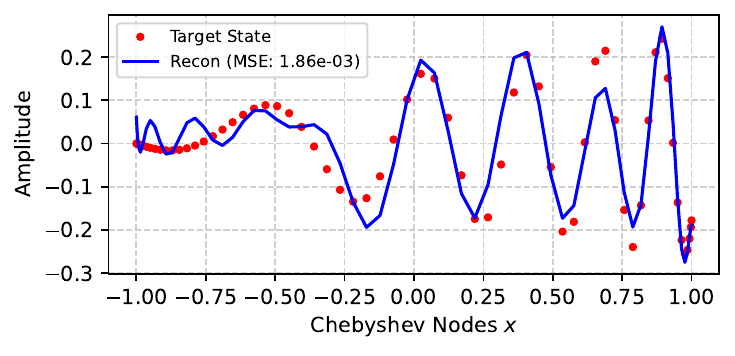}
\caption{Chebyshev Polynomial Reconstruction of $f(x) = \ln(x+2)\sin(5e^{x+1})$ encoded in $n=6$ problem qubits (64 points) using $m =20$ Chebyshev polynomials.} 
\label{fig:7}
\end{figure}

Next, we reproduced previous results validating the Hadamard test and Chebyshev polynomials as an effective method of producing an approximation of the data encoded in a quantum state with a time complexity of only $\mathcal{O}(m^d)$, using $m$ polynomials over $d$ dimensions, instead of the aforementioned exponential requirements for full statevector reconstruction. We assume an ideal method of encoding the value of $f(x) = \ln(x+2)\sin(5e^{x+1})$ at 64 points encoded in $n=6$ qubits. Then, we used 300 shots of the Hadamard test to estimate the coefficient of each of the $m=20$ Chebyshev polynomials utilized. The points chosen are Chebyshev nodes, over which Chebyshev polynomials are orthogonal, so the Gram correction matrix can be ignored. As Figure \ref{fig:7} displays, even just 20 polynomials capture the main features of a complicated function, with a mean squared error (MSE) of only $0.00186$.

\subsection{Full Quantum Subroutine to Solve Navier-Stokes for Lid-Driven Cavity}

Finally, we incorporated both of these results to perform a hybrid quantum-classical simulation of the 2D lid-driven cavity problem using $Re = 100$. The domain was discretized on a grid of $N_{grid} = 16 \times 16$ interior points ($18 \times 18$ total). The HHL solver was configured with 8 clock qubits, $n_c = 8$, and $R = 150$ Trotter steps for the Hamiltonian simulation. In each time step, the quantum state $\ket{p}$ was projected onto $m = 10$ Chebyshev polynomials. Each coefficient is calculated with a classical simulation of the Hadamard test, and subsequently adjusted via the Gram correction matrix to ensure that all $m$ polynomials are orthogonal. We note that this correction requires inverting a matrix of rank $m$, which, by conventional methods, would make this stage of the algorithm scale as $\mathcal{O}(m^3)$. Our findings indicate that even relatively small $m$ captures the fluid development with low error, so this is often negligible in the complexity of the full algorithm. To limit the runtime of these simulations, we classically emulate QAE to determine the correct scaling factor of the normalized statevector resultant of HHL. We accomplish this by extracting the factor using Qiskit's \textsf{getstatevector()} function and adding randomly generated noise to reproduce the accuracy of QAE. In particular, we introduce noise that keeps the scaling factor within $1\%$ accuracy, which would require a register of 9 qubits. Also, since the pressure and velocity are close to 0 at the bottom of the cavity, we measure error using the Normalized ARE (NARE):

\begin{equation}
    \text{NARE}(x,y) = \frac{|q_{\text{classical}}(x,y) - q_{\text{quantum}}(x,y)|}{\max |q_{\text{classical}}(x,y)|}
    \label{eq:normalized_are}
\end{equation}

\begin{figure}[h]
\centering
    \includegraphics[width=1.0\textwidth]{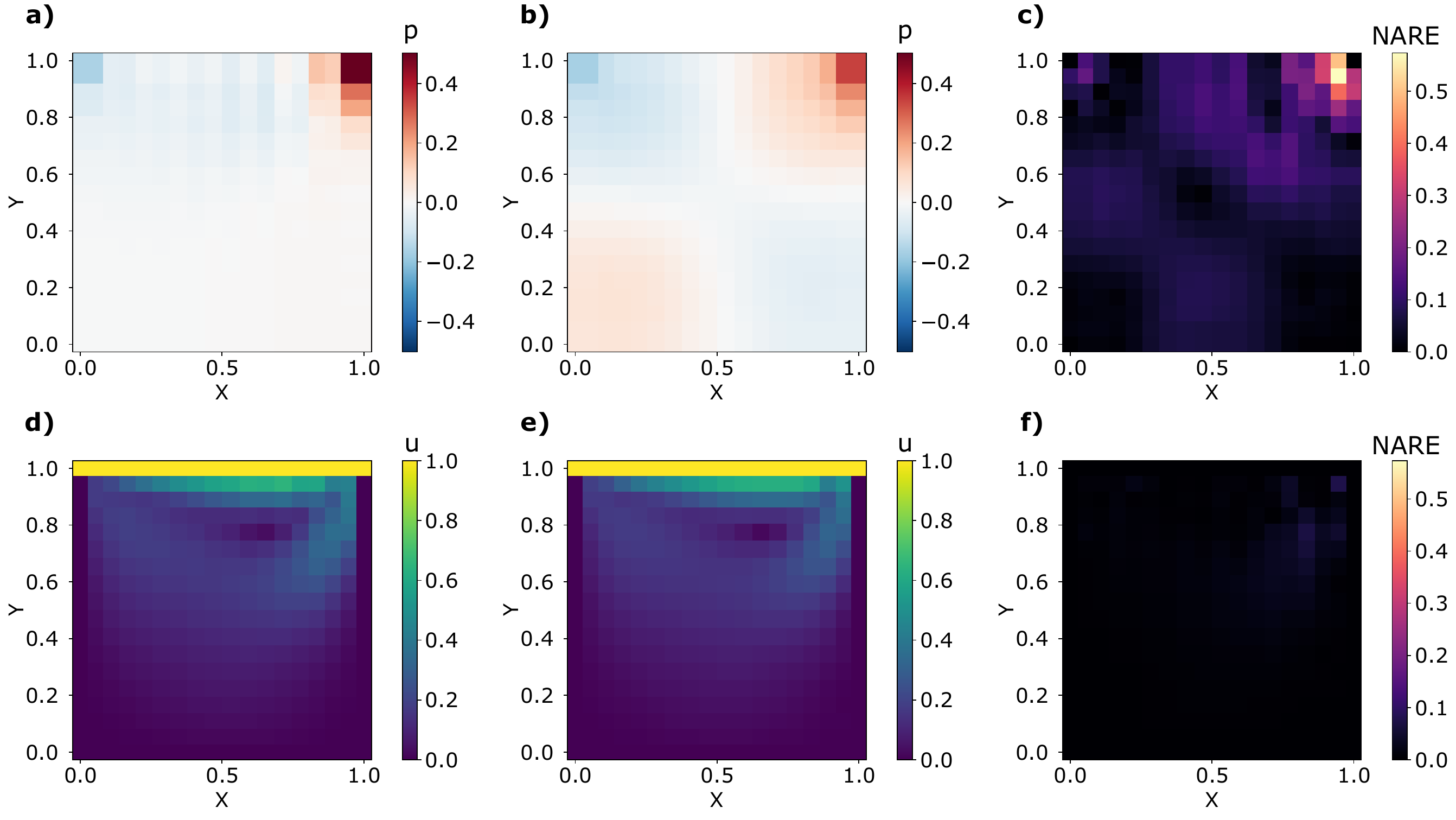}
    \caption{Comparison of 2D Lid-Driven Cavity flow ($16 \times 16$ grid), 2000 iterations corresponding to 2 seconds of fluid development. HHL Quantum-Classical Hybrid Solution with Chebyshev Polynomial QST Readout, $m=10$ polynomials, $n_c = 8$ clock qubits, and QAE register of 9 qubits. Pressure Plot: a) Classical Benchmark. b) HHL Quantum-Classical Hybrid Solution. Velocity Plot: d) Classical Benchmark. e) HHL Quantum-Classical Hybrid Solution. NARE: c) of Pressure Gradient. f) of Velocity.}
    \label{fig:8}
\end{figure}

We observed an average NARE of $0.00808$ ($0.808\%$) for the velocity field, as displayed in Figure \ref{fig:8}. The maximum NARE reached 0.075 ($7.5\%$). We found the average NARE of the pressure gradient to be $0.07094$ ($7.094\%$), with a maximum of $0.573$ ($57.3\%$).

\begin{figure}[H]
\centering
\includegraphics[width=1.0\textwidth]{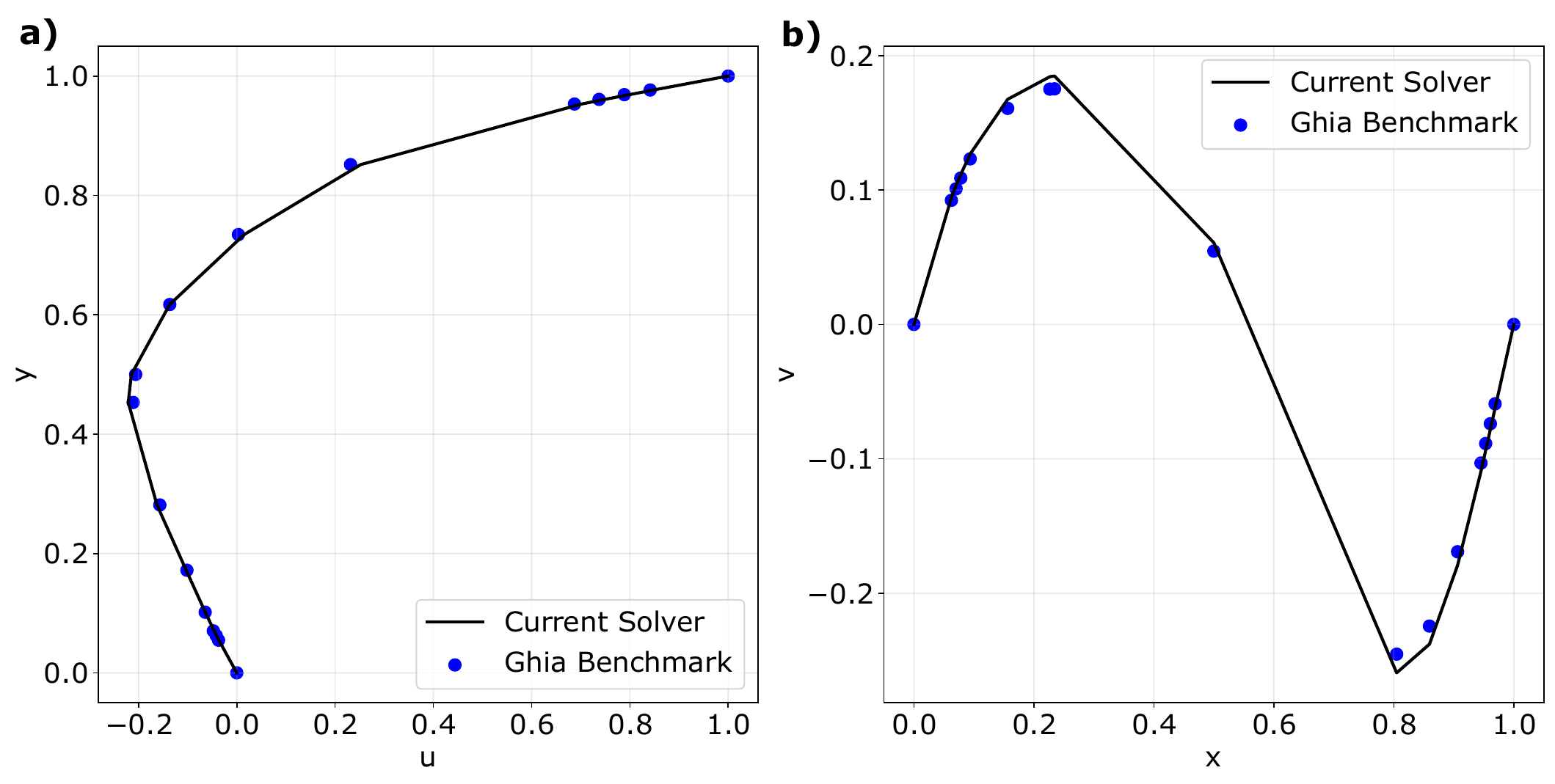}
\caption{Validation of Classical Navier Stokes Solver Against Ghia Standard on a $64\times64$ grid. a) Vertical Centerline, b) Horizontal Centerline, both comparing our classical solver to Ghia benchmark.}
\label{fig:9}
\end{figure}

\begin{figure}[H]
\centering
\includegraphics[width=1.0\textwidth]{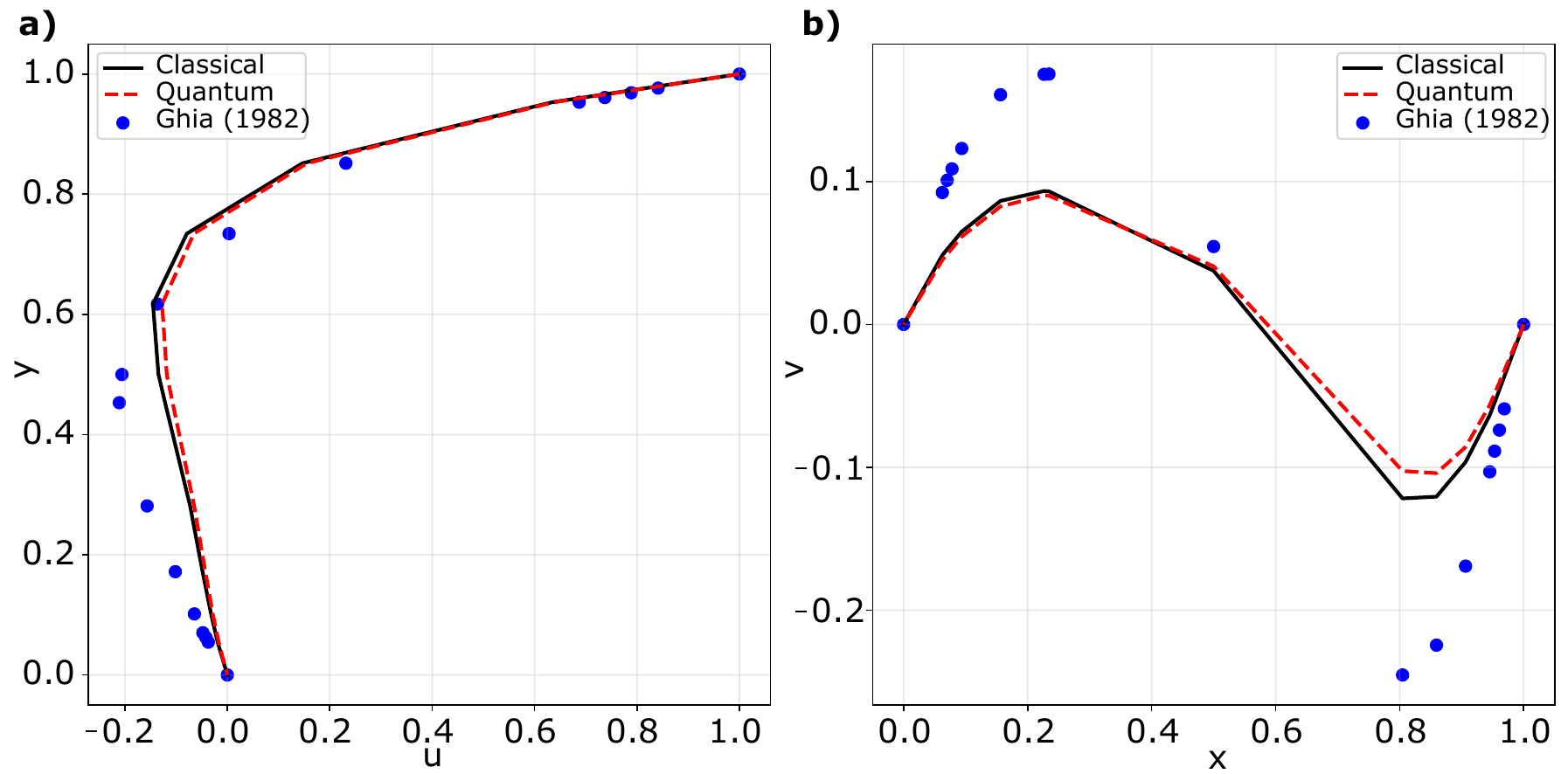}
\caption{Classical and Hybrid Quantum Solvers Centerlines Plotted with Ghia Standard, $16\times16$ grid: a) Vertical Centerline, b) Horizontal Centerline.}
\label{fig:10}
\end{figure}

To highlight the accuracy of the quantum velocity field, we also plotted both the vertical and horizontal centerlines, along with the Ghia benchmark. Both the accuracy of the quantum solver and its convergence in the direction of the Ghia benchmark are displayed in Figure \ref{fig:10}. 
Additionally, we validated our classical solver against the benchmark set by Ghia et al \cite{ghia1982high}, presented in Figure \ref{fig:9}.

\subsection{Full Quantum Subroutine to Solve Navier-Stokes for Taylor-Green Vortex}

To rigorously validate the quantum-hybrid pressure solver, we simulate the two-dimensional Taylor-Green vortex (TGV). The TGV is a classic computational fluid dynamics benchmark that models the exponential decay of a periodic array of vortices driven by viscous dissipation. Unlike the lid-driven cavity, the TGV possesses an exact time-dependent analytical solution to the incompressible Navier-Stokes equations. This allows for precise quantification of the error introduced by the quantum linear solver.

The simulation is defined on a uniform Cartesian grid over a square domain $[0, 2\pi]^2$ with periodic boundary conditions enforced across all boundaries. The exact analytical solutions for the velocity components $u, v$ and the kinematic pressure $p$ at any given spatial coordinate $(x,y)$ and time $t$ are given by:
\begin{align}
    u(x, y, t) &= \sin(x) \cos(y) e^{-2\nu t}, \label{eq:tgv_u} \\
    v(x, y, t) &= -\cos(x) \sin(y) e^{-2\nu t}, \label{eq:tgv_v} \\
    p(x, y, t) &= \frac{1}{4} \left( \cos(2x) + \cos(2y) \right) e^{-4\nu t}, \label{eq:tgv_p}
\end{align}
where $\nu$ is the kinematic viscosity of the fluid. 

The kinematic viscosity is fixed at $\nu = 0.01$ (i.e, $Re = 100$). The computational grid is initialized at $t=0$ using the exact analytical fields evaluated at the initial time step. The flow is then evolved through time using the quantum-hybrid projection method, and the terminal fields are compared against equations \ref{eq:tgv_u}-\ref{eq:tgv_p} to evaluate the solver's global accuracy.

\begin{figure}[H]
\centering
\includegraphics[width=1.0\textwidth]{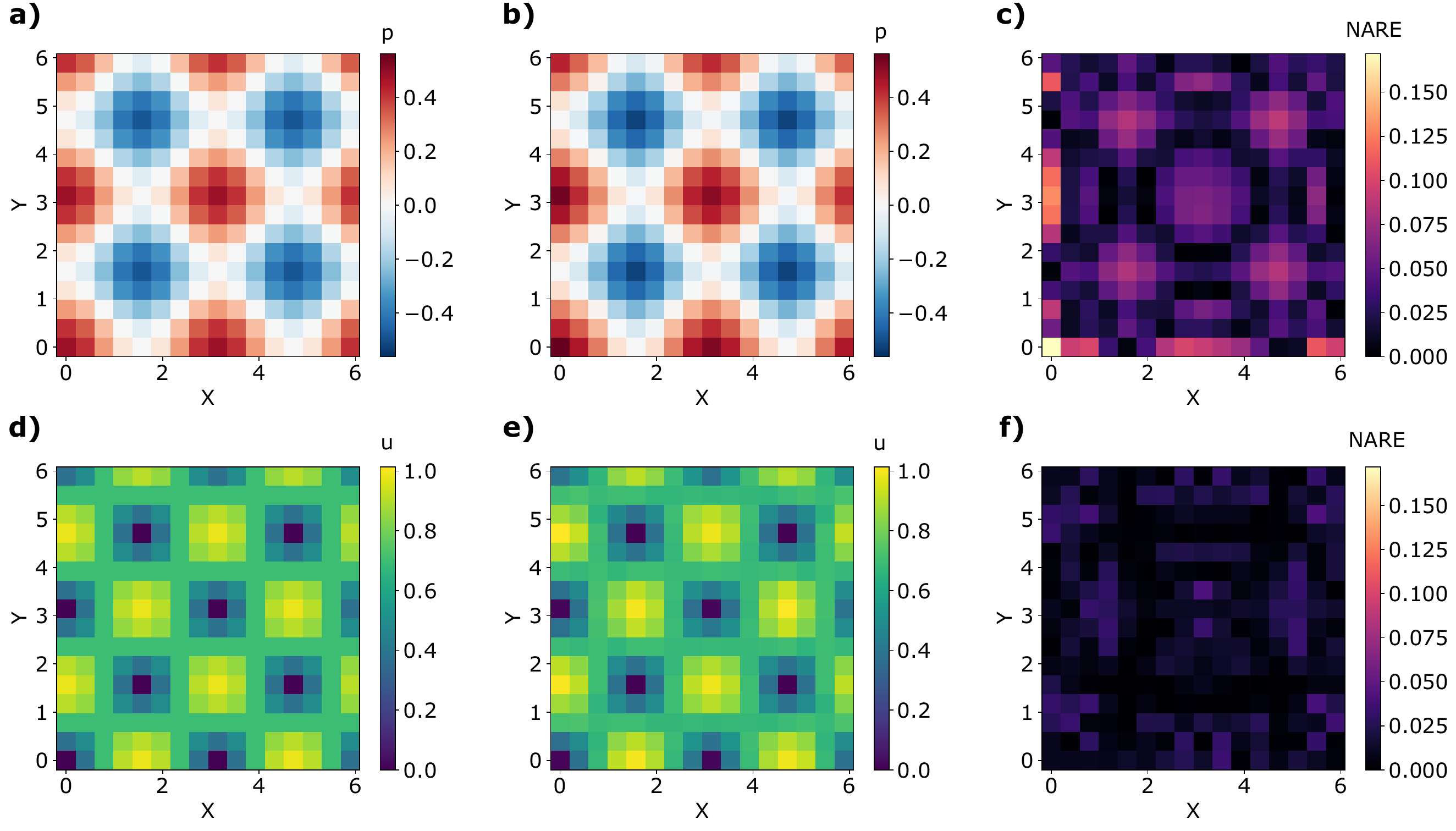}
\caption{Taylor Green Vortex on $16\times16$ Grid, Analytical Solution and HHL Quantum-Classical Hybrid Solver using $n= 8$ problem qubits and $n_c = 8$ clock qubits, readout performed with $m = 10$ polynomials, and QAE performed on 9 qubits. Pressure Plots: a) Analytical Solution, b) Hybrid Quantum-Classical Solution. Velocity Plots: d) Analytical Solution, e) HHL Quantum Classical Hybrid Solution. NARE: c) of Pressure, f) of Velocity.}
\label{fig:11}
\end{figure}

As Figure \ref{fig:11} displays, the results are very accurate, with an average NARE of only 0.0124 ($\approx1\%$) for the velocity and 0.0371 ($\approx4\%$) for the pressure. 

\section{Discussion}
The results demonstrate that the HHL algorithm can be effectively integrated into a classical operator-splitting framework for CFD. The hybrid solver maintained stability over time, proving that the quantum subroutine's noise did not lead to catastrophic divergence in the nonlinear advection steps.

Regarding the specific methodologies employed, we encountered a number of challenges in developing our quantum subroutine. Firstly, we found an inherent tension between the conditions that HHL requires to be accurate and the conditions our QST method requires to be accurate. Previous literature showed that a lower value of $\beta$, i.e, more stretched coordinates, improves the accuracy of Chebyshev QST \cite{su2025efficient}. However, HHL is very sensitive to the magnitude of the maximum eigenvalue of the matrix being inverted; the higher the maximum eigenvalue, the more clock qubits required to adequately capture the accurate solution. Decreasing $\beta$ increases the maximum eigenvalue of the Laplacian matrix, which we found to greatly affect accuracy. We found $\beta = 2.5$ to be the best balance given the number of qubits simulated. 

Additionally, we note the high error in the pressure and pressure gradient, particularly at the bottom of the cavity in the lid-driven cavity simulation (Figure \ref{fig:6}), where the pressure is close to 0. This, we believe, is the result of our QST method. A finite series of continuous functions frequently exhibits spectral "ringing," characterized by persistent oscillations within a specific band rather than converging to a stable, near-zero constant. To adequately capture both the high magnitude of pressure at the top of the cavity and the near-zero pressure at the bottom, we theorize that many more polynomials would be required. Additionally, we anticipate that for longer simulations of fluid flow, where the pressure distributes into the entire cavity, this large error would diminish significantly.

We found using the central difference method to be much more similar to the classical simulation than analytically calculating the derivative of the sum of Chebyshev polynomials. This is interesting, as an apparent advantage of this approach is the ability to analytically calculate the gradient. We suspect that analytically calculating the gradient is actually more accurate than the central difference gradient, and the distinction comes from the coarseness of the grid. We expect that the difference between these two methods would converge to 0 as the grid size increases. 

We also note the open challenge of the state preparation problem, preserving the exponential speedup quantum computers offer despite seemingly depending on operations that require access or interaction with every data point in the encoded quantum state. As mentioned above (subsection~\ref{subsec:quantum_interface}), we assume access to an ideal state preparation oracle that can encode all $N$ data points as quantum amplitudes. We hope to implement methods of encoding the initial state that preserve the theoretical exponential speedup.

\section{Conclusion}
We have presented an iterative hybrid quantum-classical solver for the incompressible Navier-Stokes equations. Although current quantum hardware limitations necessitate simulation, this work establishes a baseline for fault-tolerant quantum CFD. We offer a number of directions for future work. Firstly, the natural next step is to extend this method to the 3D Navier-Stokes equations to consider real world applications. Additionally, we seek to move beyond our idealized preparation of the initial state. Initial state preparation is one of the biggest open challenges in quantum computing, and we seek to incorporate an efficient state preparation algorithm into our quantum subroutine. Some current possibilities for this are a modified version of HHL which can efficiently encode certain types of data into a quantum state \cite{he2025solving}, or utilizing tensor networks (TN) and matrix product states (MPS) to efficiently represent and encode many types of data into quantum registers \cite{melnikov2023quantum}. Additionally, we aim to transition from fault-tolerant algorithms like HHL to VQLS. VQLS offers a more resilient pathway for NISQ hardware by replacing deep quantum circuits with a hybrid optimization loop, which naturally fits our existing iterative framework \cite{liu2023variational}. 

Alternatively, we are investigating fully quantum schemes that replace the classical advection-diffusion steps with Hamiltonian evolution to employ alongside HHL. Any quantum-classical hybrid scheme is limited by the classical computer, therefore not leveraging the broader Hilbert space for manipulating data afforded by quantum computers, nor taking full advantage of the exponential speedup, as transferring $N$ data points to and from the classical computer takes $\mathcal{O}(N)$ steps. By mapping the nonlinear Navier-Stokes equations onto a linear many-body Hamiltonian, we could potentially achieve end-to-end quantum speedup for the entire time-stepping procedure \cite{toyoizumi2024hamiltonian}, to fully leverage all the theoretical advantages quantum computers offer.
\section*{Acknowledgments}
This work was supported by the Qernel grant from the Israel Council for Higher Education.

%Bibliography
\bibliographystyle{unsrt}  
\bibliography{references}  

\end{document}